\documentclass[prl, showpacs]{revtex4}
\usepackage{epsfig}
\begin{document}

\title{High-energy photoemission on Fe$_3$O$_4$: Small polaron physics and the Verwey transition}
%------------------------------------------------------------------------------------------------------------------------
\author{D. Schrupp}
\email{david.schrupp@physik.uni-augsburg.de}
\author{M. Sing}
\author{R. Claessen}
\affiliation{Experimentalphysik II, Universit\"at Augsburg,
D-86135 Augsburg, Germany}
\author{M. Tsunekawa}
\author{H. Fujiwra}
\author{S. Kasai}
\author{A. Sekiyama}
\author{S. Suga}
\affiliation{Division of Material Physics, Graduate School of
Engineering Science, Osaka University, Toyonaka, Osaka 560-8531,
Japan}
\author{T. Muro}
\affiliation{Japan Synchrotron Radiation Research Institute,
SPring-8, Sayo-gun, Hyogo 679-5198, Japan}
\author{V.A.M. Brabers}
\affiliation{Department of Physics, Eindhoven University of
Technology, 5600 MB Eindhoven, The Netherlands}
%------------------------------------------------------------------------------------------------------------------------
%------------------------------------------------------------------------------------------------------------------------
\begin{abstract}
We have studied the electronic structure and charge ordering
(Verwey) transition of magnetite (Fe$_3$O$_4$) by soft x-ray
photoemission. Due to the enhanced probing depth and the use of
different surface preparations we are able to distinguish surface
and volume effects in the spectra. The pseudogap behavior of the
intrinsic spectra and its temperature dependence give evidence for
the existence of strongly bound small polarons consistent with
both dc and optical conductivity. Together with other recent
structural and theoretical results our findings support a picture
in which the Verwey transition contains elements of a cooperative
Jahn-Teller effect, stabilized by local Coulomb interaction.
\end{abstract}
%------------------------------------------------------------------------------------------------------------------------
\maketitle

%%%%%%%%%%%%%%%%%%%%%%%%%%%%%%%%%%%%%%%%%%%%%%%%%%%%%%%
%
% INTRODUCTION
%
%%%%%%%%%%%%%%%%%%%%%%%%%%%%%%%%%%%%%%%%%%%%%%%%%%%%%%%

Magnetite (Fe$_3$O$_4$) is not only the oldest magnetic material
known to mankind, with high potential for applications in
spin-electronics, but also displays a rather unique electronic
phase transition whose explanation has remained a challenge to
modern condensed matter physics \cite{Imada98,Walz02}. At low
temperatures magnetite is an insulator. Upon heating its
resistivity drops abruptly in a first-order phase transition at
$T_V \approx 123$~K by about two orders of magnitude to that of a
bad metal, although its temperature dependence up to 300~K remains
non-metallic. Verwey \cite{Verwey39} was the first to suggest that
the transition is related to charge ordering (CO) in this
mixed-valent oxide. In this picture the Fe$^{3+}$ and Fe$^{2+}$
ions, which occupy the octahedral $B$-sites of the inverse spinel
structure in equal numbers, form below $T_V$ an ordered pattern,
whereas above the transition they become statistically distributed
over the $B$-sublattice (with the possible preservation of
short-range order). The driving force for the Verwey transition
has been suggested to be the {\it inter}atomic Coulomb interaction
\cite{Mott90}. However, as first pointed out by Ihle and Lorenz
\cite{Ihle85}, strong coupling to the lattice may also be of
importance.

While the Verwey transition has thus been studied for decades,
very recent experiments have raised fundamental questions. Based
on new high-resolution powder diffraction data it has been
estimated that the actual charge disproportionation is only very
small, of order $\pm 0.1 e$ \cite{Wright01}. From resonant x-ray
diffraction other authors even concluded on the complete absence
of any CO in the insulating phase of magnetite
\cite{Garcia01,Garcia04}. These results suggest that the
electrostatic energy minimization implied by the proposed CO
cannot be the dominant force behind the Verwey transition contrary
to previous ideas.

Another controversial question concerns the electronic nature of
the high-temperature phase. In a high-resolution photoemission
study Chainani {\it et al.} \cite{Chainani95} found above $T_V$
finite spectral weight at the chemical potential and concluded on
a metallic state. In contrast, Park {\it et al.} \cite{Park97}
reported photoemission spectra measured just below and above $T_V$
which display a notable change in the gap but leave it finite
above $T_V$. The Verwey transition was thus identified as
insulator-to-insulator transition. We note however that all
previous photoemission experiments have been performed at low
photon energies, where the probing depth \cite{probingdepth} is
small ($\sim 15$~\AA) and the spectra are very surface-sensitive.

%%%%%%%%%%%%%%%%%%%%%%%%%%%%%%%%%%%%%%%%%%%%%%%%
%
% SHORT SUMMARY
%
%%%%%%%%%%%%%%%%%%%%%%%%%%%%%%%%%%%%%%%%%%%%%%%%

In this letter we demonstrate that low-energy photoemission
spectra of magnetite are indeed strongly affected by surface
orientation and preparation, thus limiting their reliability for a
study of the bulk electronic structure. We show how this problem
can be overcome by use of soft x-ray photoemission which due to
its enhanced probing depth ($\sim 45$~\AA) allows much better
access to the volume. It is observed that the intrinsic spectra
both above and below $T_V$ show an exponential suppression of
spectral weight towards the chemical potential and bear no
resemblance to the prominent doublet of quasiparticle peak and
lower Hubbard band as observed in other oxides with Mott-like
metal-insulator transitions. In fact, from the temperature
dependence of the spectra we conclude on a
semiconductor-semiconductor character of the Verwey transition. We
analyze the pseudogap-like line shape and demonstrate that it can
consistently be attributed to strong electron-phonon coupling. The
results yield evidence that small polaron physics plays an
essential role for the electronic properties and the Verwey
transition of magnetite.

%%%%%%%%%%%%%%%%%%%%%%%%%%%%%%%%%%%%%%%%%%%%%%%%%%%%
%
% EXPERIMENTAL DETAILS AND SURFACE PREPARATION
%
%%%%%%%%%%%%%%%%%%%%%%%%%%%%%%%%%%%%%%%%%%%%%%%%%%%%

The experiments were performed on high-quality synthetic magnetite
crystals. From these we cut oriented (111) and (100) surfaces
which after polishing were exposed to an {\it in situ} treatment
involving Ar ion sputtering and annealing (up to $800^{\circ}$~C).
Loss of surface oxygen was compensated by a subsequent oxidation
step (further details are described in \cite{Schrupp04a}).
Long-range order and stoichiometry of the surfaces were checked by
LEED and XPS, respectively. Alternatively, clean surfaces were
also obtained by fracturing single crystals {\it in situ} at
100~K. Because magnetite has no natural cleavage plane, the
resulting surfaces are rather rough and faceted. Low-energy
photoemission was performed at our home lab using He I radiation
($h\nu = 21.2$~eV). The soft x-ray experiments were performed at
BL25SU of SPring-8 \cite{Saitoh00} using Fe 2{\it p}-3{\it d}
resonance photoemission ($h\nu = 707.6$~eV). In both cases the
energy resolution was $100$~meV or lower. The position of the
chemical potential was calibrated by the Fermi edge of an
evaporated Au film in electrical contact with the magnetite
crystals.

%%%%%%%%%%%%%%%%%%%%%%%%%%%%%%%%%%%%%%%%%%%%%%%%%%%%%%%%%
%
% PRESENTATION OF DATA
%
%%%%%%%%%%%%%%%%%%%%%%%%%%%%%%%%%%%%%%%%%%%%%%%%%%%%%%%%%

\begin{figure}[ht]
\begin{center}
\epsfig{file=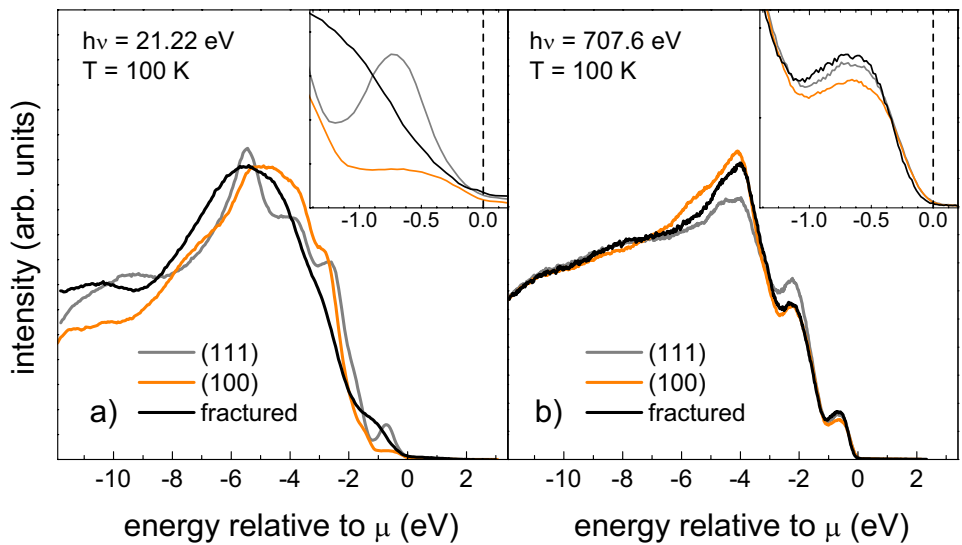,clip=,width=13cm}
\caption{\label{Figure1} (a) Valence band spectra of three
differently prepared and oriented magnetite surfaces, measured
with vacuum ultraviolet radiation. Inset: Spectral weight
distribution near the chemical potential. Note the dramatic
surface effects on the low-energy electronic structure. (b)
Corresponding spectra measured with soft x-ray radiation at the Fe
L$_2$ edge. The strongly reduced differences between the various
surfaces give evidence of the enhanced bulk sensitivity.}
\end{center}
\end{figure}

Figure \ref{Figure1}(a) and (b) show the valence band spectra of
the different magnetite surfaces measured with vacuum ultraviolet
(VUV) and soft x-ray (SX) radiation, respectively. The VUV spectra
display dramatic differences depending on surface preparation and
orientation, reflecting strong electronic structure rearrangements
in the outermost atomic layers. In contrast, the SX spectra show
only very small (though reproducible) differences, indicating that
due to the larger probing depth they are dominated by signal from
the bulk. This is particularly evident in the spectral weight
distribution near the chemical potential $\mu$ (see insets in
Fig.~\ref{Figure1}) which originates from the $d^6 \rightarrow
d^5$ transition at the {\it B}-site Fe$^{2+}$ ions
\cite{Anisimov96,Yanase99} and is directly related to the
low-energy physics of magnetite. For all three surfaces the SX
data yield the same spectral line shape with a broad maximum at
$\approx -0.6$ eV and a strong suppression of intensity towards
the chemical potential. We hence conclude that this is the
intrinsic low-energy spectrum of magnetite.

%%%%%%%%%%%%%%%%%%%%%%%%%%%%%%%%%%%%%%%%%%%%%%%%%%%%%%%%
%
% TEMPERATURE DEPENDENT SPECTRA OF THE FRACTURED SAMPLE
%
%%%%%%%%%%%%%%%%%%%%%%%%%%%%%%%%%%%%%%%%%%%%%%%%%%%%%%%%

\begin{figure}
\begin{center}
\epsfig{file=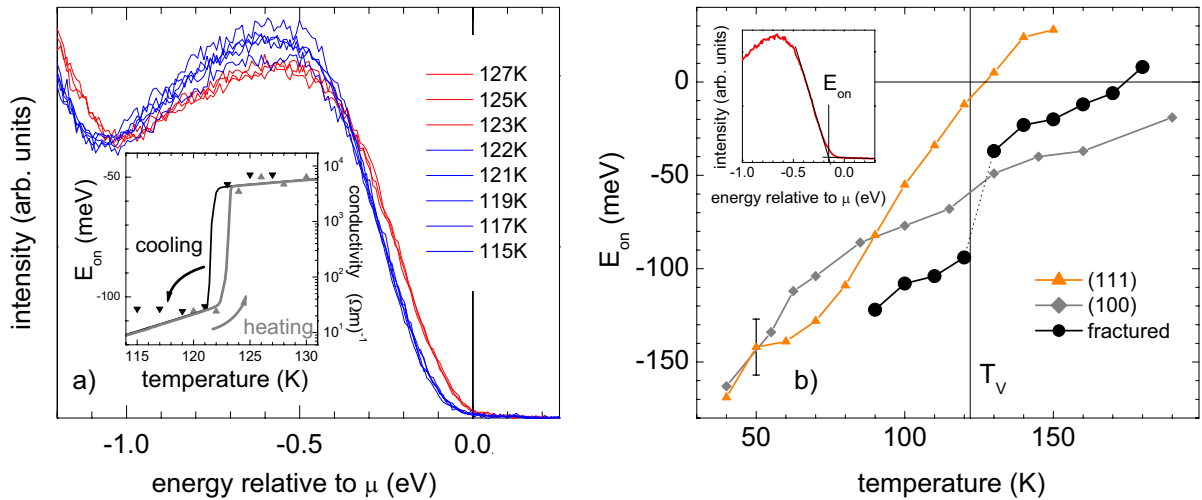,clip=,width=16cm}
\caption{\label{Figure2}(a) Spectra of the fractured sample taken
around the Verwey transition. Inset: spectral onset energy during
cooling (down-triangles) and heating (up-triangles). Also shown is
the conductivity hysteresis (solid curves) measured on the same
sample. (b) Temperature-dependence of the spectral onset energy
(see inset for definition) over a wider temperature range (see
inset for definition). Also shown are the results for the annealed
(111) and (100) surfaces.}
\end{center}
\end{figure}

For a further analysis of the $d^6 \rightarrow d^5$ signal we now
turn to its temperature dependence. Figure \ref{Figure2}(a) shows
SX spectra of the fractured surface measured for temperatures
within a few Kelvin around $T_V$. Below the Verwey transition the
spectrum is consistent with a small energy gap, as expected for an
insulator. Going through the Verwey transition the spectral onset
becomes abruptly shifted towards the chemical potential. However,
no indication of a metallic Fermi edge or quasiparticle feature is
observed in the high-temperature phase, in notable contrast to the
spectra of oxide materials with a Mott transition such as, e.g.,
V$_2$O$_3$ \cite{Mo03}. For a more quantitative examination we
define the spectral onset phenomenologically as the intersection
of the leading edge with the zero intensity base line (see inset
of Fig.~\ref{Figure2}(b)). A plot of this phenomenological
parameter versus temperature is shown in the inset of
Fig.~\ref{Figure2}(a). The onset energy jumps exactly at $T_V$ and
is consistent with the hysteretic behavior of the conductivity,
confirming that these spectra reflect intrinsic bulk behavior.
This conclusion is further corroborated by the observation that
the onset energy agrees well with the activation energy of the
conductivity on both sides of the transition \cite{Walz02}. If the
onset energy is indeed identified with the insulator half-gap
(with $\mu$ assumed to be in mid-gap position), its discontinuous
change $\Delta E_{on} \approx 50$~meV will cause the concentration
of thermally activated charge carriers to increase by a factor of
$\exp (\Delta E_{on}/k_B T_V )$. This accounts quantitatively for
the observed two-order of magnitude jump in the conductivity, as
was already argued by Park {\it et al.} \cite{Park97}. A decrease
rather than a complete closing of the gap at $T_V$ also identifies
the Verwey transition as an insulator-to-insulator transition
\cite{Park97}.

We also note that temperature effects are not restricted to the
immediate vicinity of $T_V$. The discontinuity observed at the
Verwey transition is actually superimposed on top of a continuous
energy shift of the spectral onset over a wide temperature range,
as shown in Fig.~\ref{Figure2}(b). The identification of a
first-order Verwey transition thus requires a very fine
temperature grid around $T_V$. Previous photoemission studies,
besides being strongly surface-sensitive, used rather wide
temperature steps, so that the reported effects resulted (at least
partly) from the gradual temperature evolution and not from the
Verwey transition.

%%%%%%%%%%%%%%%%%%%%%%%%%%%%%%%%%%%%%%%%%%%%%%%%%%%%%%%
%
% SURFACE EFFECTS
%
%%%%%%%%%%%%%%%%%%%%%%%%%%%%%%%%%%%%%%%%%%%%%%%%%%%%%%%

Interestingly, the discontinuous first-order transition of the
onset is only observed for fractured surfaces, but not for the
(111) and (100) surfaces (also shown in Fig.~\ref{Figure2}(b)).
For the (111) sample the onset shifts continuously towards the
chemical potential and reaches it at $\sim T_V$ without any
evidence for a discontinuous phase transition or a metallic Fermi
edge. The onset of the (100) spectra shows also a continuous but
much smaller temperature shift. Moreover, the onset remains
clearly below the chemical potential even up to 190~K, well above
$T_V$. It is tempting to relate this finding to reports of
room-temperature charge order at the (100) surface
\cite{Wiesendanger92, Shvets04}. These surface effects in the
temperature evolution can be traced back to the fact that any
termination of a bulk magnetite crystal will result in a polar
surface, which in order to minimize its energy has to undergo a
geometric (or electronic) relaxation. The resulting atomic
reconfiguration may then lead to a suppression of the Verwey
transition, similar to the adverse effects of other lattice
deformations in the volume induced by defects or hydrostatic
pressure \cite{Walz02,Rozenberg96,Brabers99}. For the (111) and
(100) samples the thickness of the relaxed surface layer is
obviously comparable to the relatively large photoemission probing
depth and therefore extends over at least one unit cell (lattice
constant $a=8.4$~\AA). As their preparation involves high
temperature annealing, the resulting surface configurations are in
thermal equilibrium. In contrast, crystal fracture at cryogenic
temperatures will result in a non-equilibrium surface in which the
topmost atomic layer may be disturbed but a full surface
relaxation is hindered by the lack of sufficient thermal energy.
The subsequent layers (accessible by SX photoemission) will remain
frozen in a bulk-like configuration and thus display the Verwey
transition.

%%%%%%%%%%%%%%%%%%%%%%%%%%%%%%%%%%%%%%%%%%%%%%%%%%
%
% POLARON INTERPRETATION OF BROAD LINE SHAPE
%
%%%%%%%%%%%%%%%%%%%%%%%%%%%%%%%%%%%%%%%%%%%%%%%%%%

\begin{figure}
\begin{center}
\epsfig{file=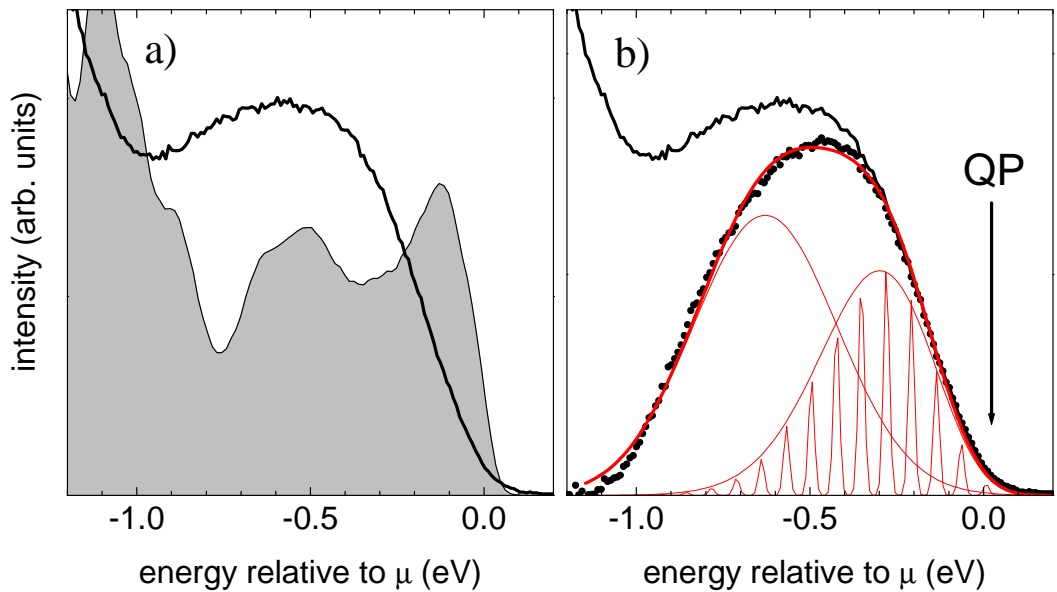,clip=,width=14cm}
\caption{\label{Figure3} (a): Near-$\mu$ spectrum of a fractured
Fe$_3$O$_4$ crystal at 180~K ({\it i.e.}~above $T_V$) superimposed
on the band theory DOS for the high-temperature phase
\cite{Yanase99}, cut-off by a Fermi function for the same
temperature. (b): Model fit (red curve) of the
background-subtracted spectrum (black dots) and its decomposition
into the polaronic single-particle spectrum of
Ref.~\cite{Alexandrov92} and an additional Gaussian (see text for
details). "QP" denotes the location of the exponentially
suppressed polaronic quasiparticle band.}
\end{center}
\end{figure}

In any case, the above results establish that the broad spectral
distribution and the absence of a well-defined quasiparticle peak
near the chemical potential are intrinsic properties of the
electron removal spectrum and not a surface artefact. The
non-observation of a sharp Fermi edge, particularly above $T_V$,
is in clear disagreement with LDA band calculations
\cite{Yanase99} which predict a metallic state (see
Fig.~\ref{Figure3}(a)). Furthermore, even if one allows for an
insulating state as suggested by the activated conductivity on
both sides of the transition, the inflexion point of the leading
edge of the spectrum (which in a band insulator would mark the gap
edge) lies at much higher binding energy than expected from the
transport (half)gap. Rather, the gap coincides with the energy of
the spectral onset (as defined above), very reminiscent of
observations recently made in photoemission spectra of
quasi-one-dimensional Peierls systems and attributed there to
polaronic effects \cite{Perfetti01,Perfetti02}. In this picture
strong electron-lattice coupling leads to the formation of
polaronic quasiparticles, {\it i.e.} electron (or hole)
excitations heavily dressed by virtual phonons. Removing an
electron from the coupled system (as in photoemission) results in
a spectrum consisting of a coherent quasiparticle peak, greatly
reduced in spectral weight and renormalized in energy, and an
incoherent background of phonon side bands shifting weight away
from the chemical potential. If the coupling is sufficiently
strong, a pseudogap-like behavior of the spectrum will result.
Indeed, for magnetite strong coupling to the lattice and the
formation of small polarons have already been inferred from other
properties, e.g., the small carrier mobility \cite{Tsuda91}, the
unusual temperature behavior of the conductivity above $T_V$, and
the observation of a mid-infrared polaron peak in the optical
conductivity \cite{Park98,Gasparov00}, from which a polaronic
binding energy of $\varepsilon_p \sim 300$~meV has been derived.

In order to further substantiate the above qualitative picture we
have analyzed the data using the theoretical model of Alexandrov
and Ranninger \cite{Alexandrov92}, in which the electron removal
spectrum of a system with strong electron-phonon coupling can be
written as
\begin{equation}
I(\varepsilon) \propto e^{-g^2} \tilde{N}_p (\varepsilon) +
\sum_{n=1}^{\infty} e^{-g^2}\frac{g^{2n}}{n!} \tilde{N}_p
(\varepsilon + n\omega_0).
\end{equation}
The first term represents the polaronic quasiparticle band
($\tilde{N}_p (\varepsilon)$) renormalized in spectral weight and
width ({\it i.e.}~its inverse mass) by a factor $e^{-g^2}$, where
$g^2 = \varepsilon_p/\omega_0$ is a dimensionless electron-phonon
coupling constant and $\omega_0$ a characteristic phonon energy.
The remaining intensity is transferred to multiple phonon side
bands shifted by $n\omega_0$ away from $\mu$. For strong coupling
($g^2 >> 1$) the polaron quasiparticle weight is thus
exponentially suppressed in the photoemission spectrum, but its
binding energy still defines the lower edge of the transport gap.
According to band theory the (non-coupling) density of states
(DOS) near the chemical potential consists of two peaks, one right
at $\mu$ and a second one at $-0.6$~eV (Fig.~\ref{Figure3}(a)).
The polaronic effects in the electron removal spectrum have now
been modeled for the first DOS peak by eq.~(1) and for the second
by a Gaussian of similar width \cite{Gaussian}.
Figure~\ref{Figure3}(b) shows the result for the 180~K spectrum of
the fractured sample, with the phonon energy scale $\omega_0$
fixed to the highest optical mode (70 meV \cite{Gasparov00}) and
the theoretical spectrum broadened by $\sim 100$~meV to account
for instrumental resolution and a possible residual polaron
bandwidth. The best agreement with experiment is obtained for a
coupling strength of $g^2 \sim 5$, which accurately reproduces the
observed pseudogap behavior near the chemical potential and the
shift of spectral weight to higher binding energy with respect to
the band theory DOS. The value for $g^2$ is consistent with both
the polaronic binding energy and the effective polaron mass ($\sim
100\dots200 m_0$) derived from infrared spectroscopy
\cite{Gasparov00}. This analysis corroborates the qualitative
arguments from above for a pronounced electron-phonon coupling and
formation of small polarons in magnetite.

%%%%%%%%%%%%%%%%%%%%%%%%%%%%%%%%%%%%%%%%%%%%%%%%%%
%
% JT EFFECT AS ORIGIN OF STRONG COUPLING
%
%%%%%%%%%%%%%%%%%%%%%%%%%%%%%%%%%%%%%%%%%%%%%%%%%%

The strong coupling to the lattice can be traced back to the fact
that Fe$^{2+}$ in a high-spin $d^6$ configuration is a Jahn-Teller
(JT) ion. In fact, the low-temperature distortions of the oxygen
octahedra surrounding the formally divalent B-site ions can be
understood as JT-active modes \cite{Wright01}. Very recent LSDA+U
calculations \cite{Leonov04,Jeng04} based on the refined structure
of Ref.~\cite{Wright01} confirm this view. They find that the
local occupation of $t_{2g}$-orbitals on the $B$-sublattice
follows octahedral distortions as expected from the JT effect,
except that the actual charge disproportionation is less than $\pm
0.1e$, leading to non-integer Fe valencies ($2.4+$/$2.6+$). As a
consequence, the low-temperature phase is characterized by orbital
order (OO) and much less so (if at all) by CO, consistent with the
recent diffraction experiments \cite{Wright01,Garcia01,Garcia04}.
It is thus tempting to conclude that the physics of the Verwey
transition contains elements of a cooperative JT effect, which
requires additional stabilization by local $dd$ Coulomb
interaction \cite{Leonov04,Jeng04}.

%%%%%%%%%%%%%%%%%%%%%%%%%%%%%%%%%%%%%%%%%%%%%%%%%%%%
%
% CONCLUSION
%
%%%%%%%%%%%%%%%%%%%%%%%%%%%%%%%%%%%%%%%%%%%%%%%%%%%%

We thus arrive at a picture in which the charge carriers in
magnetite are small polarons, presumably of the JT type, forming a
narrow band of high effective mass. At high temperatures they
contribute to the charge transport by both coherent tunneling and
activated hopping, as originally proposed by Ihle and Lorenz
\cite{Ihle85}. Below $T_V$ the polarons condense into an
orbitally-ordered phase (possibly with an additional small CO
amplitude), whose microscopic origin remains to be revealed. This
opens a gap in the polaronic quasiparticle band and causes the
sudden drop in the conductivity. The polarons are much more stable
than the OO/CO as inferred from our spectra of the annealed
surfaces, for which the Verwey transition is suppressed but not
the polaronic line shape. Our results demonstrate that the Verwey
transition in magnetite is not a purely electronic effect and that
any microscopic model has to account for the important role of
electron-phonon coupling in the simultaneous presence of
interatomic Coulomb interaction.

\acknowledgments We thank A.S.~Alexandrov, H.~Fehske, S.~Fratini,
T.~Kopp, I.~Leonov, A.~Loidl, A.~Pimenov, and D.~Vollhardt for
helpful discussions. The photoemission experiments were performed
with approval of the Japan Synchrotron Radiation Research
Institute (Proposal Nos. 2002B0104-NS1-np and 2003B0159-NSa-np).
Funding was provided by the Deutsche Forschungsgemeinschaft
through SFB 484. M.S. gratefully acknowledges financial support by
the Japan Society for the Promotion of Science (JSPS).


\begin{thebibliography}{0}
\bibitem{probingdepth}
Probing depth is defined as three times the inelastic mean free
path corresponding to 95\% of the photoemission signal.
\bibitem{Gaussian}
A Gaussian profile is the limiting shape of eq.~(1) for strong
coupling. Here we are interested in shape and position of the
pseudogap behavior near $\mu$ which is dominated by the polaron
effects in the first DOS peak. Therefore a simple Gaussian
approximation for the second peak is fully sufficient.
\end{thebibliography}
\end{document}